\documentclass[showpacs,secnumarabic,amssymb,nobibnotes,aps,prl,11pt]{revtex4-2}
\usepackage{amsmath}
\usepackage{amssymb}
\usepackage{graphicx}
\usepackage{tabularx}
\usepackage{gensymb}
\usepackage{setspace}
\usepackage{url,hyperref}

\usepackage[usenames,dvipsnames]{xcolor}
\hypersetup{colorlinks=true, linkcolor=BrickRed, urlcolor=blue!50!black, citecolor=blue!50!black}
\usepackage[capitalize]{cleveref}
\usepackage{cleveref}
\AtBeginDocument{}
\crefrangelabelformat{equation}{(#3#1#4)$-$(#5#2#6)}
\begin{document}

\title{Simulations of Mpemba Effect in WATER, Lennard-Jones and Ising Models: Metastability vs Critical Fluctuations}

\author{Soumik Ghosh}
\author{Purnendu Pathak}
\author{Sohini Chatterjee}
\author{Subir K. Das}
\email{das@jncasr.ac.in}
\affiliation{Theoretical Sciences Unit and School of Advanced Materials, Jawaharlal Nehru Centre for Advanced 
Scientific Research, Jakkur P.O., Bangalore 560064, India}
\date{\today}

\begin{abstract}

Via molecular dynamics simulations we study ICE formation in the TIP4P/Ice model that is known
to describe structure and dynamics in various phases of WATER accurately. For this purpose well equilibrated
configurations from different initial temperatures, $T_s$, belonging to the fluid phase, are quenched
to a fixed subzero temperature. Our results on kinetics, for a wide range of $T_s$, following such quenches, show quicker
crystallization of samples that are hotter at the beginning. This implies the presence of the puzzling Mpemba effect (ME). Via a similar study,
we also identify ME in fluid to solid transitions in a Lennard-Jones (LJ) model. In the latter
case, the ME appears purely as an outcome of the influence of critical fluctuations on the nonequilibrium growth process, for which we present interesting scaling results. For the TIP4P/Ice case, on the other
hand, we show that delay in nucleation, due to metastability, can alone be a driving factor for
the exhibition of ME. To substantiate the difference between the two cases, we also present LJ-like
scaling results for ME in a magnetic transition. Our simulations indicate that in each of
the systems the effect can be observed independent of the cooling rate that may vary 
when samples from different $T_s$ are brought in contact with a heat reservoir
working at a fixed lower temperature.

\end{abstract}

\maketitle

\section{Introduction}

A hotter sample of Water may freeze faster, than a colder one, when kept inside a refrigerator working at a 
subzero temperature \cite{bechhoefer2021, bechhoefer2020, skdLang2023, jeng2006}. This counterintuitive fact is referred to as the Mpemba Effect (ME) \cite{mpemba1969} and is 
discussed since the time of Aristotle \cite{arristotle2005}. In recent times, in efforts to generalize the effect, 
several other experimental systems and theoretical models were shown to exhibit similar phenomena.  Examples include
cooling granular gases \cite{santos2017, rajesh2020}, clathrate hydrates \cite{ahn2016}, anti-ferromagnets \cite{raz2017} and spin glasses\cite{parisi2019}, as
well as, perhaps even more surprisingly, pure ferromagnetic systems \cite{skdLang2023, vadakkayil2021, chatterjee2024}. Despite the continuously growing list of such systems, the ME remains still a puzzle. There exists no clear hint whether a common fact is responsible for the observations in different systems. In fact, for the original system, i.e., Water, even a demonstration via computer simulations is nonexistent, though there are works by providing possible explanations if the effect indeed exists \cite{auerbach1995, goddard2015, zhang2014, gijon2019, bechhoefer2020, bechhoefer2021}. 
Such a status is perhaps due to the difficulty owing to the complex natures of Water molecules and
related interactions \cite{carrasco2009} that do not allow simulations of adequately large systems for long enough times. The long simulations are needed to counter the
metastable features that may severely delay nucleation for a transition from a fluid phase to Ice. An important question here is to ask: How the ME may be connected to the metastability -- Should the longevity of the latter be a function of the initial temperature? Or, growth, following the nucleation, is
the only contributor to the initial temperature dependence of the transformation? Questions have
also been raised if the ME is a result of differences in times for reaching the final temperature, $T_f$, from different starting temperatures, $T_s$ \cite{auerbach1995, Vynnycky2015}. To
address these issues, and obtain a more general picture, in addition to theoretical and
computational studies of Water, it should also be of interest to undertake studies of simpler systems exhibiting fluid to solid transitions. 
Here, we study such transitions in a model Water \cite{matsumoto2002, abascal2005_2} and a Lennard-Jones (LJ) system \cite{allenbook}, without impurities. 

In each of the considered cases, we prepare equilibrium fluid configurations at different $T_s$. 
These configurations are quenched to a fixed $T_f$, at which the thermodynamic phase is a solid one. In the case of Water, achieving Ice nucleation under \textit{homogeneous}
condition is a difficult task, for reason already mentioned above. We choose a model \cite{abascal2005_2} for which this was demonstrated \cite{matsumoto2002} to occur with reasonable ease. 
Note that higher possibility of fluctuations should enhance the scope of nucleation. Consideration of small systems, may severely restrict such chances. 
On the other hand, handling large systems is a difficult task, when the requirement is to simulate for long times that may be a necessity for the present problem for a certain
range of $T_s$. Thus, it is important to adopt a ``trade-off'' between system size and run length \cite{skdLang2023}. 
Following this strategy, we have been able to realize homogeneous Ice nucleations for a large range of $T_s$. From 
the analysis of the corresponding data sets we find clear evidence of ME in this as well as the LJ system. 
In the case of LJ, metastability is not a matter of concern; expected thermodynamic structure for the chosen $T_f$ is always obtained without encountering any barrier. This and the
results for Water provide interesting classification of ME based on the role of metastability. We also present results for para-to-ferromagnetic transitions in a model system
\cite{chatterjee2024, vadakkayil2021, landaubook}, Hamiltonian of which does not contain any element of
frustration. This is to further substantiate the conclusion from the LJ system that ME can be observed in absence of
metastability as well \cite{vadakkayil2021, skdLang2023, arnab2023}, unlike the case of Water, driven, for example, by the differences in critical fluctuations at various starting points \cite{vadakkayil2021, skdLang2023, chatterjee2024}.

\section{models and techniques}

For the LJ system, we perform molecular dynamics (MD) simulations \cite{allenbook, frankelbook}
using a truncated, shifted, and force-corrected potential, within which two particles at a distance $r$
apart interact via \cite{allenbook, skd2006}
$u(r)=U(r)-U(r_c)-(r-r_c)(dU/dr)_{r=r_c}$. Here $U(r)=4\varepsilon\left[ (\frac{\sigma}{r})^{12}-(\frac{\sigma}{r})^{6}\right]$ is the standard LJ potential, $\sigma$ being the particle diameter and 
$\epsilon$ deciding the strength of the interaction. The cut-off distance is taken to be \cite{skd2006} $r_c=2.5\sigma$. The modified form captures the same basic facts of transition (e.g., critical universality remains unchanged) as the original one, while speeding up the simulations. In $d=2$, the space 
dimension of our interest for this model, corresponding coexistence curve in the number density ($\rho$)-temperature ($T$) plane was
estimated using Monte Carlo (MC) simulations \cite{midya2017}. The critical values of $\rho$ and $T$ were noted to be $\rho_c =0.37$, and 
$T_c=0.41\varepsilon/k_B$, respectively, $k_B$ being the Boltzmann constant. While one observes \textit{practically} a vapor-liquid coexistence for $T$ close to $T_c$, far below $T_c$ the high density phase is a solid one, in
$d=2$ this being a (quasi long-range) hexatic one. For this model, MD simulations are carried
out in constant NVT ensemble, N and V being, respectively, the number of particles and volume of the system, using a hydrodynamics preserving Nos\'{e}-Hoover (NH) thermostat
\cite{nose1990, frankelbook}, while quenching homogeneous
configurations from different $T_s$ $(>T_c)$ to a $T_f$ belonging to the coexistence regime. We have
used square boxes of linear dimension $L=256\sigma$. The integration time step was fixed at
$\Delta t=0.005\tau$, $\tau$ $(=\sqrt{m\sigma^2/\varepsilon})$ being the LJ unit of time,
and $m$ the mass of each particle. We set $\varepsilon$, $\sigma$, $k_B$ and $m$ to unity.

For the simulations of Water, we have chosen a rather realistic model \cite{abascal2005, matsumoto2002}: \textbf{TIP4P/Ice}. 
This is a four-point rigid Water model with transferable intermolecular potential. There, in 
addition to the points related to the positions of one oxygen (O) atom and two hydrogen (H) 
atoms, an additional point exists. This point, having a (negative) charge $1.1794$, in 
electronic unit, is placed at a fixed distance ($0.1577 \AA$) from O along the bisector of the
H-O-H angle, having experimental value \cite{abascal2005_2} $104.52\degree$. The O atoms interact with
each other via the full LJ potential, with $\varepsilon=0.21084$ 
kcal/mole. Each H-atom carries a positive charge of magnitude $0.5897$. All charge points,
expectedly, interact via the Coulombic potential \cite{abascal2005_2}. The masses of O and H atoms are set to be
$15.9994$ amu and $1.008$ amu, respectively. With this model we carry out NPT MD
simulations, with $96$ molecules \cite{Chen2017, Piaggi2020, Buch1998}, using LAMMPS \cite{LAMMPS}, having NH thermo- \cite{nose1990, frankelbook} and barostats \cite{martyna1994}, at pressure $P=4.5$ atm, for which the thermodynamic 
limit boiling and freezing points stand \cite{Antonio2006} at $\simeq420K$ and $\simeq273K$, respectively. Each
simulation we run upto $500$ ns, to observe freezing, with the integration time step $\Delta t=1$ fs. Results on average freezing time are obtained by using the runs that exhibit freezing \cite{moore2011, matsumoto2002}.

For the magnetic case, we carry out MC simulations with the Ising model, having the Hamiltonian
\cite{landaubook} $H=-J\sum_{<i,j>}S_iS_j$, $S_i=\pm1$ being the possible spin states and $J$ 
$(=1)$ the interaction strength, in $d=2$, for which $T_c\simeq2.269 J/k_B$.
For this model, we have implemented the Glauber spin-flip kinetics \cite{landaubook} on a square
lattice having same box dimensions as the LJ case. An MC trial
move there is to randomly choose a lattice site and change its spin state. Such a move is
accepted following a standard Metropolis criterion \cite{landaubook}. The time in these simulations are measured in units of MC steps (MCS), a unit
consisting of $L^2$ trial moves. For all the models, we have applied periodic boundary conditions. 

\begin{figure} [tb]
\centering
\includegraphics*[width=0.48\textwidth, height=0.38\textwidth]{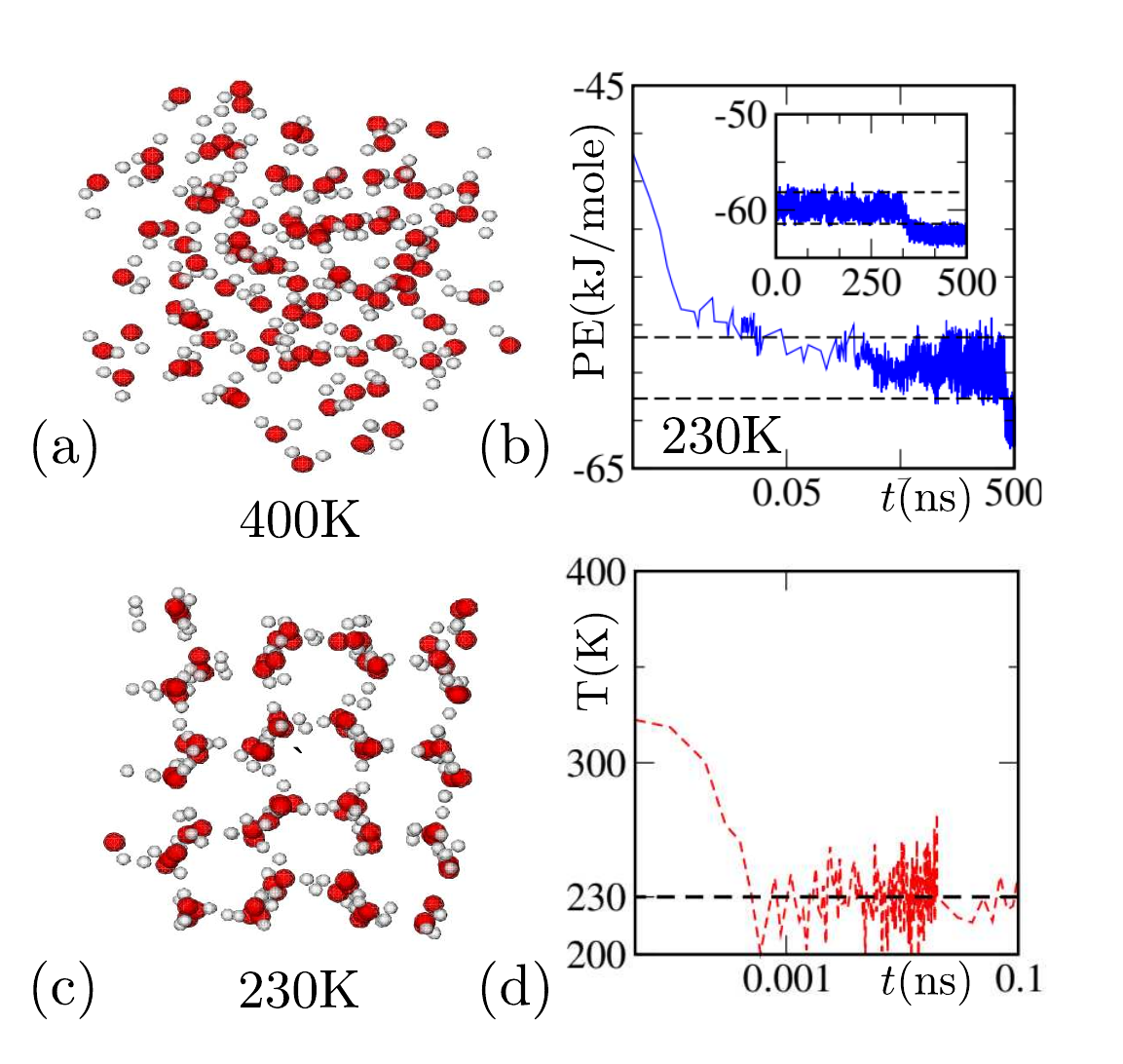}
\caption{(a) An equilibrium configuration of Water molecules is shown for $T_s=400K$. (b) Potential energy profile corresponding to a simulation run, starting with the snapshot at (a), is shown, following a quench to $T_f=230$K. In the main frame we have used a log scale for time, whereas in the 
inset the scale is linear. The jump corresponds to metastable Water to Ice transition. The horizontal lines capture the fluctuations of energy within the metastable window. (c) A typical 
configuration is shown from the Ice regime of the last simulation run. (d) Temperature is plotted
versus time for the simulation in (b). The assigned value of temperature is shown by the dashed line.}
\label{water1}
\end{figure}

\begin{figure}[tb]
\centering
\includegraphics*[width=0.48\textwidth]{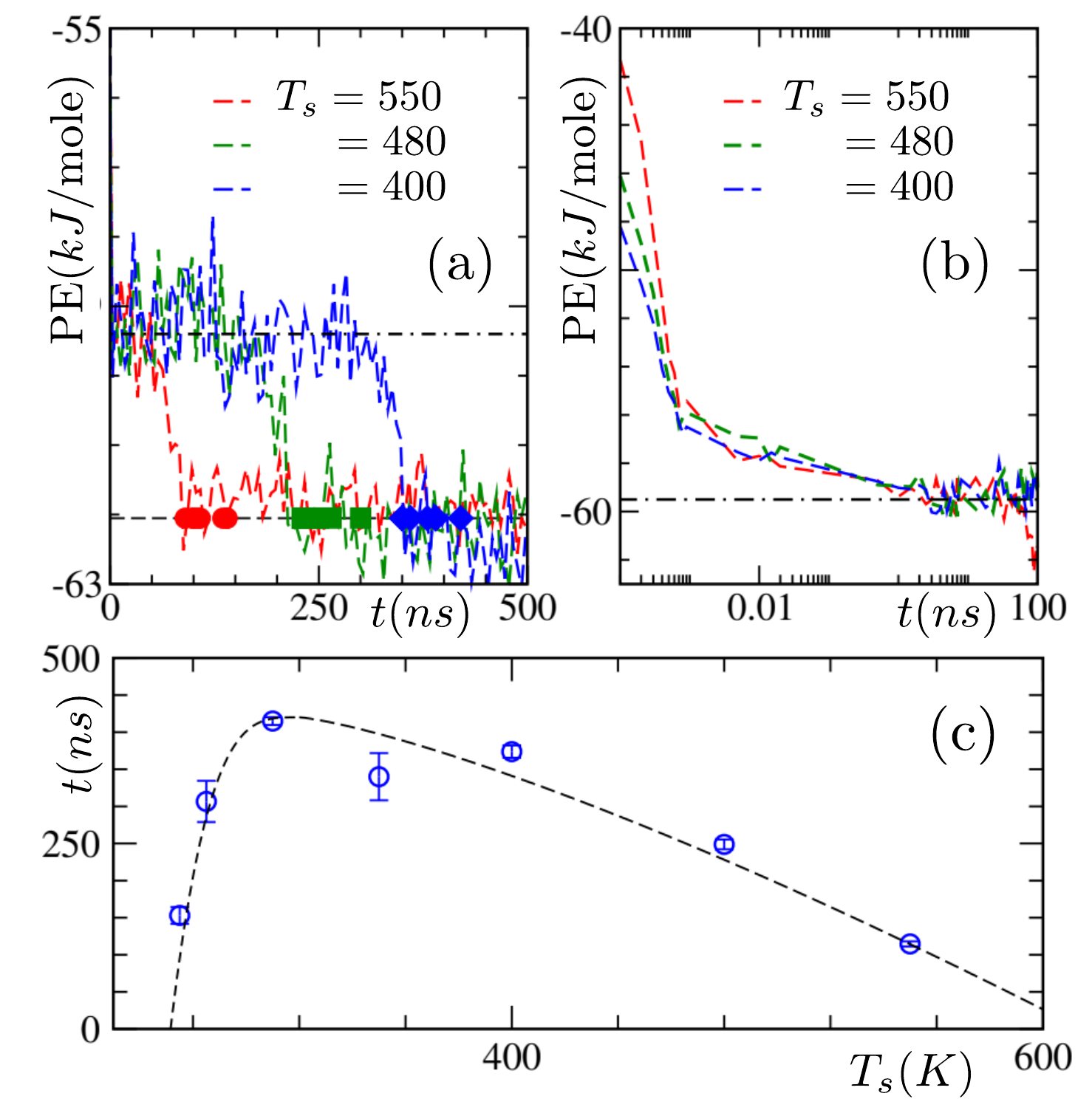}
\caption{ (a) PE is plotted as a function of time for the TIP4P/Ice simulations. Results 
are shown for quenches to $T_f=230$K, from a few $T_s$. Data in each of the sets are 
thinned down for the sake of clarity. The symbols show the locations of jumps for simulation runs with certain other starting
configurations. A unique colour is used for a given
$T_s$. The early parts of the energy decay are shown again in (b) by using a log scale for 
the abscissa. The horizontal lines stand for approximately the average values for intermediate metastable liquid and final Ice phase energies. 
(c) The freezing time, averaged over $8$ runs, for each $T_s$, is plotted versus $T_s$. The dashed line is a guide to the eye.}

\label{water2}
\end{figure}

\section{Results}

Fig. \ref{water1} provides information on the freezing of a configuration of water molecules, when
quenched from $T_s=400K$ to $T_f=230K$, following the protocol described below. A system of Water molecules were first arranged on a periodic array, with bond lengths and orientations taken from GitHub \cite{Github_water}, before heating up to $500K$ to rapidly break the periodicity. A fluid configuration from 
this heating run is further simulated at the desired $T_s$, to prepare initial configurations, before finally
quenching to $T_f$. A typical configuration (see a fluid-like snapshot in (a)) from the equilibrium part of the later run is quenched to the
previously mentioned $T_f$. The potential energy (PE) corresponding to this simulation is shown in
part (b), over $500$ ns. For a long period, the energy fluctuates
within a window corresponding to a metastable liquid phase, before jumping down to another
window that relates to an Ice phase, which we have confirmed via calculations of appropriate structural quantities. Corresponding hexagonal structure is shown in part (c). A
temperature profile for this run can be seen in part (d). Clearly, the temperature settles at
the assigned $T_f$ within a short time interval.

\begin{figure}[tb]
    \centering
    \includegraphics*[width=0.48\textwidth]{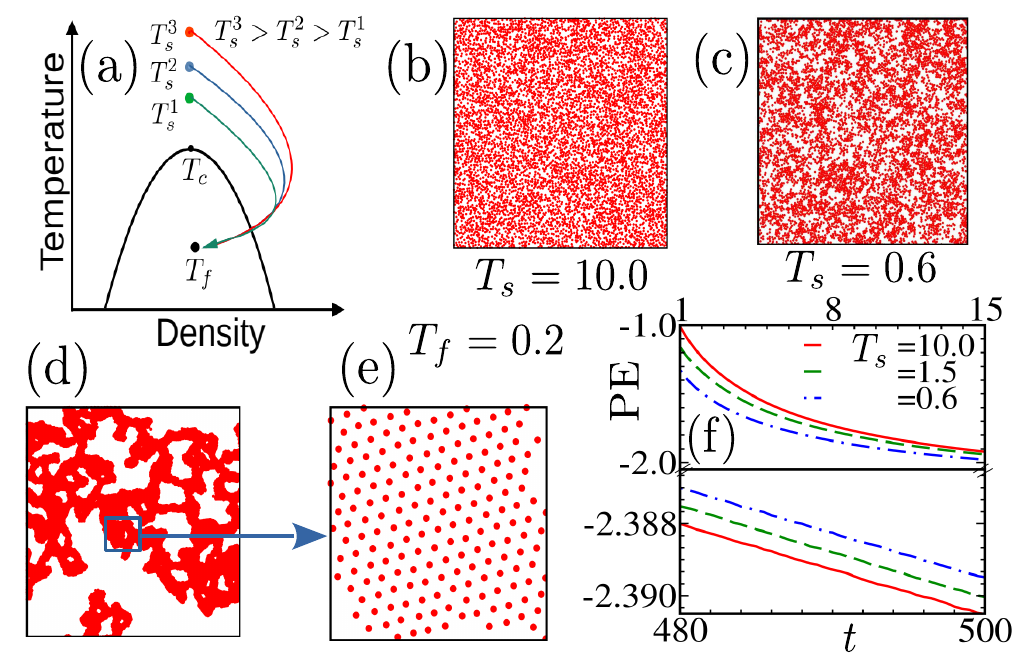}
    \caption{ (a) Schematic diagram describing the protocol for studying kinetics in the LJ model. In (b) and (c) we show typical equilibrium snapshots from
    two different $T_s$. (d) A late-time snapshot, captured during an
    evolution, following a quench from $T_s=0.6$ to $T_f=0.2$, is shown. (e) An enlarged portion of the snapshot in (d) is showing a
    hexatic arrangement of particles. (f) Plots of PE, following quenches to $T_f=0.2$, from a few $T_s$, are shown against time. The upper half displays the early-time 
    behavior and the lower one corresponds to the late-time decays.}
    \label{LJ1}
\end{figure}

Fig. \ref{water2} (a) shows the energy profiles for the earliest freezings for a set of $T_s$
values. The transformation times, from the intermediate metastable to the Ice phase, appear
systematic, for the presented range of $T_s$. For a clear visualization of the arrivals at the
metastable energy levels, certain early parts of the plots are shown in a semi-log scale in Fig. \ref{water2} (b). 
These results are indicative of faster decay of PE for higher $T_s$. This resembles ME in
other systems that we will describe below. In Fig. \ref{water2} (a), we have marked the freezing
times corresponding to simulation runs starting with some other configurations. 
For each $T_s$, the spread of the times is restricted to a rather narrow window. We present the
key results, on $t_f$, the average freezing time, in Fig. \ref{water2} (c). 
This plot strongly points towards the existence of the ME, while nicely resembling the nonmonotonic
character seen in the original work \cite{mpemba1969}, even without impurity, i.e., for homogeneous nucleation. Since in experiments heterogeneous nucleation cannot be avoided, our results discard a possible belief that ME in Water may be an outcome related to impurities.
Noting that the temperature settles at the final value rather early, see Fig. \ref{water1} (d), compared to the freezing times, differences in heat transport or cooling rate for different $T_s$ are also unlikely to have roles to play in the effect. 
Before further discussing the origin of this fascinating behavior in Water, we take a look at the results from other systems.

Fig. \ref{LJ1} corresponds to the LJ model. In part (a) we show the protocol. Systems from 
various $T_s$ are quenched to a fixed $T_f$ inside the coexistence curve. Typical initial configurations are shown in (b) and (c). 
Clearly, these snapshots differ from each other in terms of spatial fluctuations that can be
quantified via $\xi$, the equilibrium correlation length \cite{fisher1967}. Fig. \ref{LJ1} (d)
shows a typical nonequilibrium configuration obtained during an evolution following a quench to
$T_f=0.2$, from $T_s=0.6$. An enlarged portion of this snapshot is displayed in (e) which shows a regular arrangement of particles \cite{midyaprl}. In part (f) we show
the decay of potential energy for quenches from a few $T_s$. The upper and lower parts of this
frame contain, respectively, the early and late time data. Opposite sequences of appearances of the curves for
different $T_s$, in the two time regimes, indicate crossings that imply a faster rate of
equilibration for a hotter system, the basic requirement of ME \cite{parisi2019, vadakkayil2021}. Tendency of such crossings can be appreciated from early data in the case of
Water as well -- see the plots in Fig. \ref{water2} (b) -- faster rate of fall for higher $T_s$ is clear. In this case, the crossings, however, got delayed
due to the appearances of metastability.  Before discussing further the case of Water, we present more results on the LJ case, as well as similar ones from the Ising model.

\begin{figure}[tb]
    \centering
    \includegraphics*[width=0.48\textwidth]{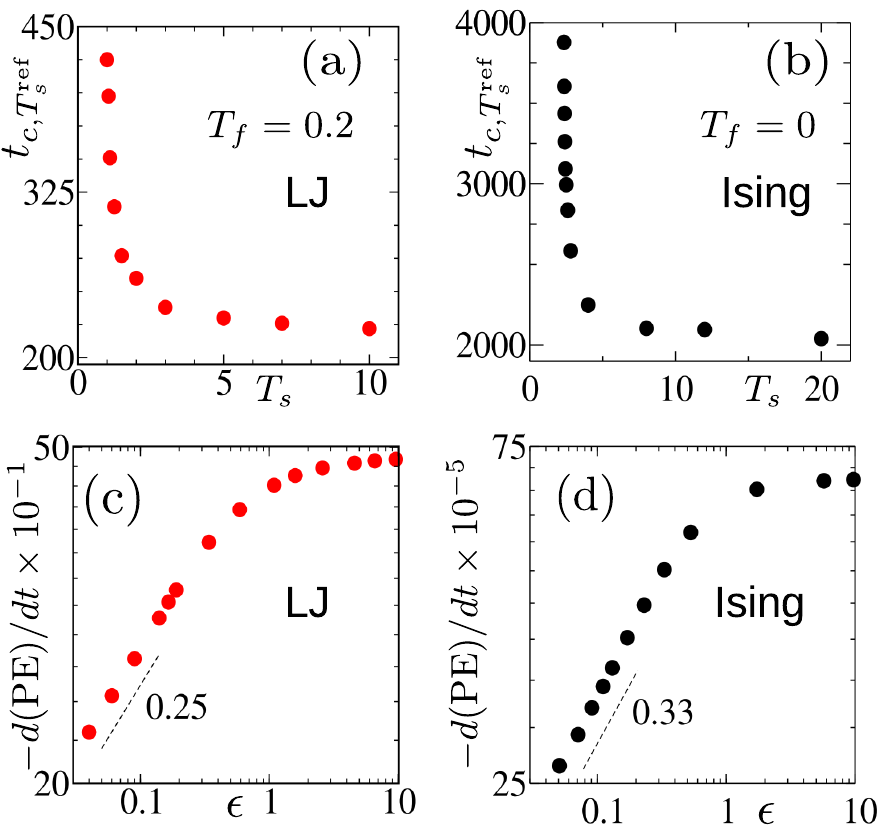}
    \caption{ (a) Plot of the crossing times, $t_{c,T_s^{\rm{ref}}}$, as a function of $T_s$, for the LJ model. (b) Same as (a), but here it is for the Ising model with $T_f=0$. The results were obtained after averaging over runs with nearly $20000$ independent initial configurations. (c) Negative of the decay rates of $\rm{PE}$, for different $T_s$, are plotted versus $\epsilon$, for the LJ model, with $t=7$. (d) Same as (c), but here it is for the Ising model, with $t=50$. The dashed lines in (c) and (d) represent power-laws.}
    \label{LJ_Ising}
\end{figure}

We have noted the times, $t_{c,T_s^{\rm{ref}}}$, corresponding to the crossings of the PE curves for
various $T_s$ with that for a reference case $T_s^{\rm{ref}}=0.6$. These are plotted in Fig. \ref{LJ_Ising} (a), as a function of $T_s$. 
In Fig. \ref{LJ_Ising} (b) we show a similar plot from the Ising model with $T_s^{\rm{ref}}=2.32$ as the reference case. Very systematic trend is visible, implying that the systems from each higher $T_s$ are approaching the new equilibrium quicker than those from any lower $T_s$. 
Flattening of the plots with the increase of $T_s$, for both the models, however, implies that for very high $T_s$ values, even if the difference is large, between two starting $T_s$, ME will be weak. Thus, the effect in these two models may be related to differences in critical fluctuations \cite{skdLang2023} at different $T_s$, that die out when  $T_s>>T_c$. 
To facilitate a quantitative understanding, we calculate $d{\rm(PE)}/dt$, the rate of decay of
$\rm{PE}$, for different $T_s$. For the ME to exist, $|d{\rm (PE)}/dt|$ should decrease with the
decrease of $T_s$, when estimated at a fixed value of $t$. If the ME is truly due to the differences in 
critical fluctuations at the initial states, we may expect a scaling behavior of this quantity,
when seen against $\epsilon=T_s-T_c$. Given that these two models belong to the same static
universality class \cite{landaubook,skd2006,midya2017,fisher1967},
this critical behavior can possibly be reasonably close, barring deviation due to some differences in the values of the domain growth exponent \cite{landaubook, midya2020} and finite-size effects between the two cases.
In Fig. \ref{LJ_Ising} (c) and (d), we demonstrate it, for specific
choices of $t$, for the LJ and the Ising cases, respectively. Indeed, asymptotically, reasonably
similar power-law behavior emerge, while indicating very slow equilibration as $T_s \rightarrow T_c$.
On the other hand, the rate is nearly constant for $T_s>>T_c$.
This overall picture supports our conjecture on the role of critical fluctuation.
To appreciate the statement on such fluctuations, it is worth looking at the snapshots in Figs. \ref{LJ1}
(b) and (c). 

In the case of Water, on the other hand, such differences in spatial correlation are practically nonexistent, for the considered initial state points. 
It should be noted that the vapour-liquid critical point in this case has the coordinates $(T_c, P_c) \equiv (647.096\rm K, 22.064 \rm MPa)$. This is far from the trajectory of our initial states.
We have, nevertheless, verified this quantitatively by calculating the density field structure factors for large enough systems and comparing with the
Ornstein-Zernike picture \cite{stanleybook}. Thus, for Water we have a different problem in
hand.

\section{Conclusion}

We have presented results from the studies of kinetics of transitions in pure Water, from fluid phases to 
that of Ice. The relaxation, quantified via the transition time, starting from various
initial temperatures, follows an interesting nonmonotonic behavior. For a large range of starting
temperatures, surprizingly, a hotter sample freezes earlier than a colder one. This nicely resembles the 
Mpemba effect \cite{mpemba1969}.
For comparison, we have also presented results from simulations of simpler systems for which
differences in critical fluctuations at the initial states appear to be the driving factor for the ME. It
is evident that the origin is different in Water. In this case, the behavior in
Fig. \ref{water2} (c) is related to the jumps from metastable intermediates to the Ice phase.
The longevity at a metastable level should be connected to the properties, associated with molecular orientations, hydrogen bonding, etc., at the corresponding initial state. 
In addition, it will also be interesting to undertake studies by combining the
distribution of nuclei in space, and of their occurrences in time, with the growth process, for large enough systems. 

\section{author contribution}

SKD proposed and designed the overall project, supervised the work and wrote the manuscript. SG carried
out all the works on Water. PP and SC contributed, respectively, the results on the LJ and the Ising
models. 

\section{acknowledgement}

The authors acknowledge computation times in the clusters of National Supercomputer
Mission located in JNCASR.

\end{document}